# Radio Frequency Modulated Signaling Interconnect for Memory-to-Processor and Processor-to-Processor Interfaces: An Overview

Jason Y. Du, Student *Member, IEEE*

***Abstract*— With the evolution of heterogeneous computing system, such as network-on-chip, high-performance distributed computing, accelerator-rich architectures and cluster computing, high-speed, energy-efficient and low-latency interfaces among memory-to-processor and processor-to-processor become the key technology to enable those technologies. Simultaneously, the scaling of CMOS makes the switching speed of the transistor up to sub-THz. Radio-frequency or even millimeter-wave modulated signaling interconnect has unique features in ultra-low power operation, dynamic allocation of bandwidth and low latency, compared with convention baseband signaling interconnect. In this work, we overview the different generations of radio-frequency interconnect (RF-I) technology, compare them with conventional baseband signaling interconnect technologies. The limitations and potentials are also discussed in the end.**

## I. INTRODUCTION

To meeting the even-increasing computation-intensive applications and the demands of low-power, low-cost and high-performance system, the number of heterogeneous computing systems in a single chip has enormously increased, such as network-on-chip (NoC), high-performance distributed computing (HPDC), accelerator-rich architectures (ARA) and cluster computing (CC) [1]. The key and common requirement of those systems are a high-speed, energy-efficient and low-latency interconnect technology, which supports communication among memory to the processor, processor to processor, accelerator to memory, accelerator to accelerator, and accelerator to processor. Microsoft's Project Catapult is a great example in hyper-scale cloud-based acceleration by utilizing conventional baseband high-speed interconnects on Altera's Stratix V D5 FPGA [2].

At the same time with the key benefits of the ultra-scaling CMOS technology, the switching speed of transistor increases a lot over each technology node. Based on ITRS reports [3], $f_T$ and $f_{max}$, will exceed 800 GHz and 1 THz, respectively in 10nm CMOS technology. With the advance of CMOS radio-frequency and millimeter-wave circuits, higher and higher bandwidth will be available shortly. Recently, many published works demonstrated millimeter-wave band more than 60 GHz [4-9] and even up to THz range [10-14]. With more frequency band resources, CMOS-based circuits are driving all kinds of radio-frequency related applications. For instance, the CMOS RF circuits are used for wireless and wireline communication [15-21], human-machine interfaces [22-24], navigation [25], etc.

Power consumption and heat dissipation are very critical issues of modern high-performance computing platform [2]. For example, Fig. 1 shows that the serial interface power consumption is almost comparable with computing core's power. Another example published by Intel, saying the serial interface power is going to exceed 50% of total CPU power with higher and higher IO data rate in the very near future.

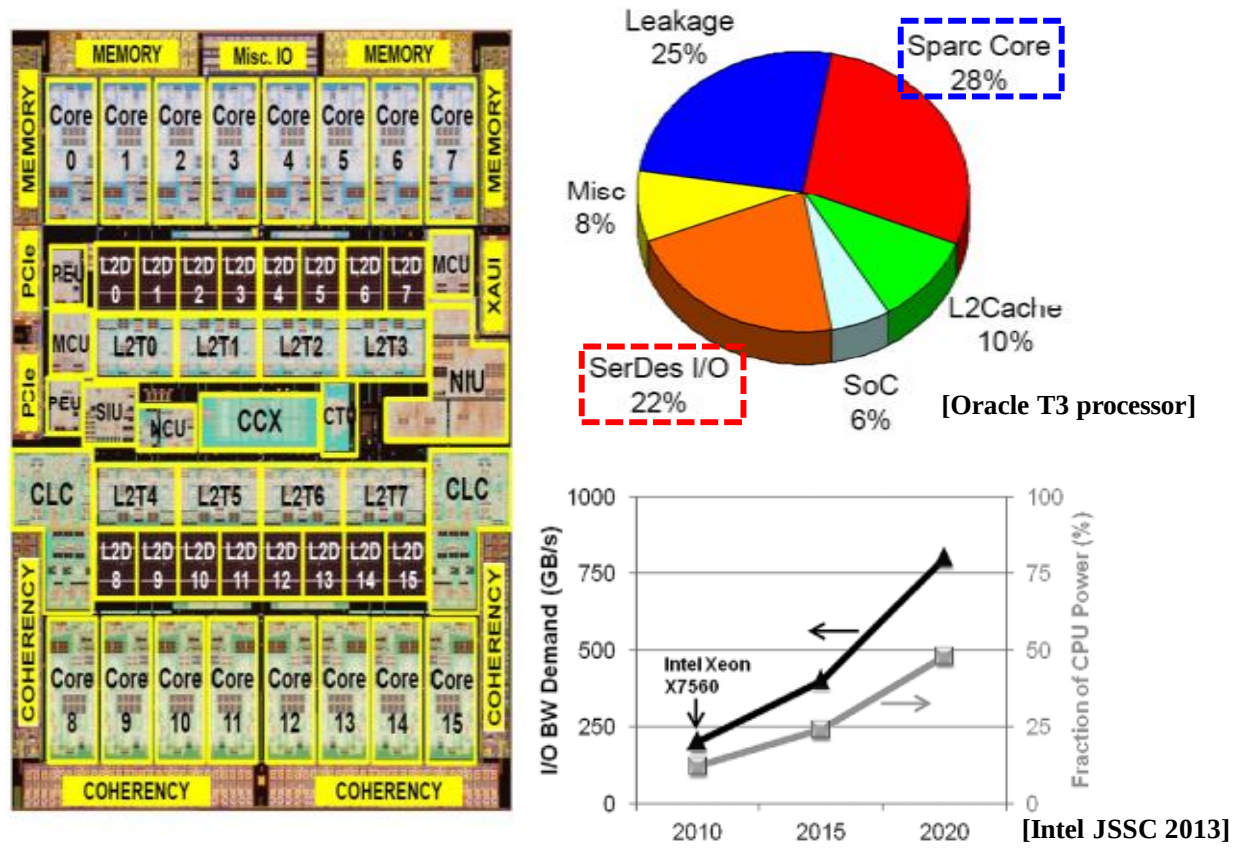

Fig. 1 Trend of power consumption of interconnect

In this paper, the conventional baseband signaling interconnects, and equalization techniques will be reviewed in Section II and III. In Section IV and V, we will introduce radio-frequency signaling concept and summarizes the radio-frequency interconnect (RF-I) technology development generation by generation. Section VI will draw the conclusion.

## II. CONVENTIONAL BASEBAND SIGNALING INTERCONNECT

The data rate of peripheral serial input/output (I/O) for PC and mobile computing platforms continue to scale to meet high-bandwidth applications including high-resolution displays, camera sensors and large-capacity external storage [26]. In Fig. 2 (a) the blue curve is technology scaling, and the red curve is the number of functions per chip, increasing dramatically with technology scaling.

More interestingly, Fig. 2 (b) shows that the CPU clock rate does not change much over 15 years, mainly because power and heat dissipation becomes a severe problem. Moreover, similarly, the IO pads number per chip also remains relatively constant but for a different reason, mainly because of packaging cost. Nowadays, the packaging plus testing have already taken more portion than fabrication in the semiconductor industry, especially for high-speed high-pin-count chips. It is very obvious that there is a large gap between IO data rate and internal data rate, 15 ~ 20 times difference.

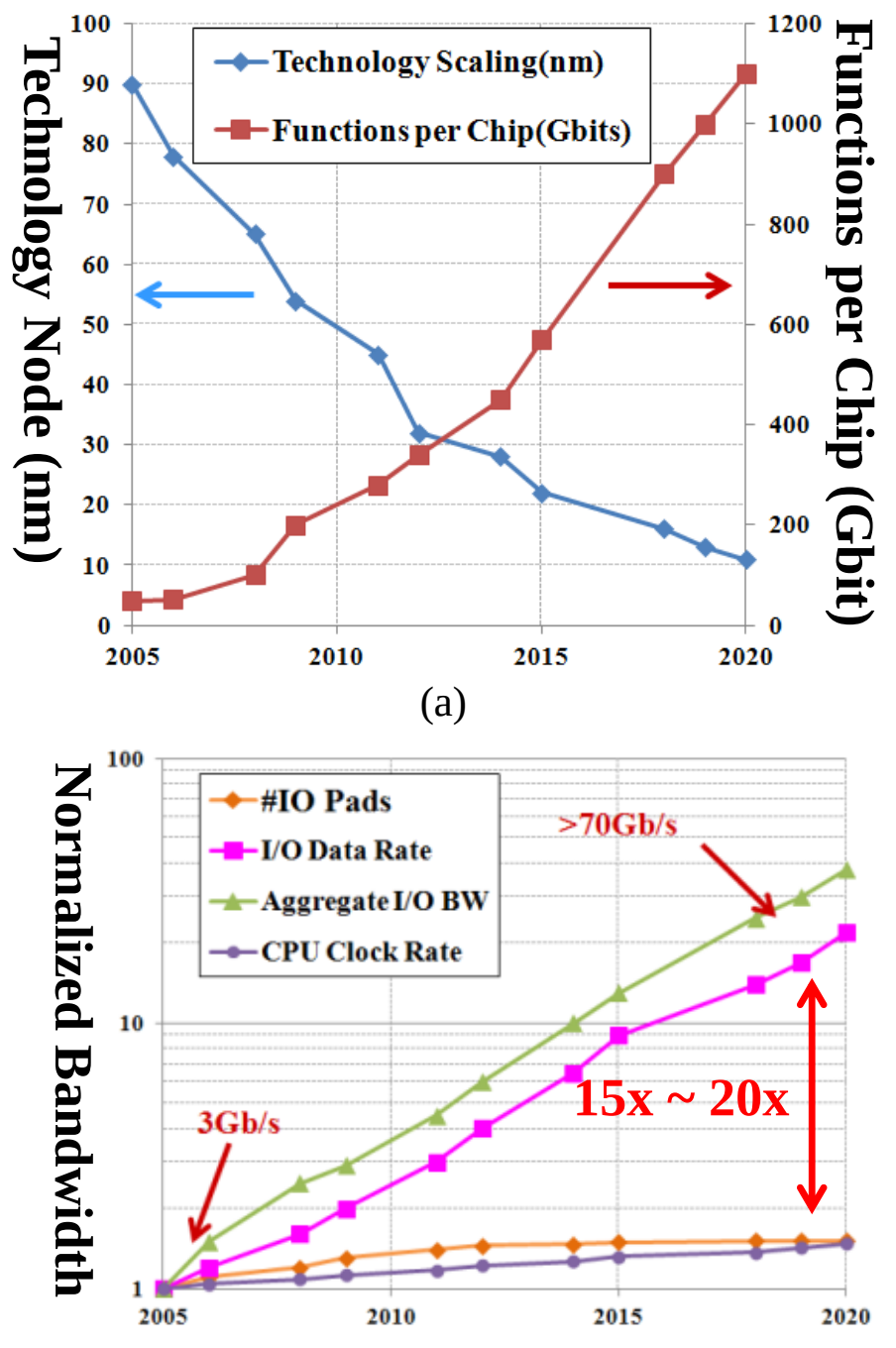

Fig. 2 (a) Technology trend with functions per chip, (b) data bandwidth, and CPU clock rate

Interconnect design evolved a lot over these years as Fig. 2 shows. Starting from 1980's, when the data rate is around tens of Mb/s, the interconnect on-chip routing wire, cable or copper traces on printed circuits board (PCB) could be modeled as a simply lumped capacitor. The transceiver design was CMOS inverter-based driver and receiver.

Then in 1990's and 2000's, data rate gets to hundreds of Mb/s and Gb/s level, the similar interconnects (on-chip routing wire, cable or copper traces on PCB) have to be modeled as a transmission line, which has more distributive effects, such as characteristic impedance mismatch, inter-symbol-interference (ISI), incident/ reflected wave superposition, and so on. At that time, a lot of parallel interfaces with the multi-drop features, like PCI, old generation DDR were replaced by point-to-point serial interfaces.

Today, the data rate of mobile or PC computing platform interfaces data rate are going to exceed more than 10 Gb/s. There are also some cables, vias, and novel substrate materials progress to improve signal integrity. However, if compared with the data rate increasing, they are far more enough to overcome the communication channel bandwidth limitation. It has to rely on complicated and power-hungry equalization technique. Moreover, transmission line channel model is not accurate enough, all the discontinuities and non-ideal effects, including vias, bumps, bonding wires, pads, traces in packages, connectors, etc., should be modeled carefully. It is tough to meet both data rate requirement and power and cost budget at the same time. It is one of the hottest areas both in academic and industrial fields.

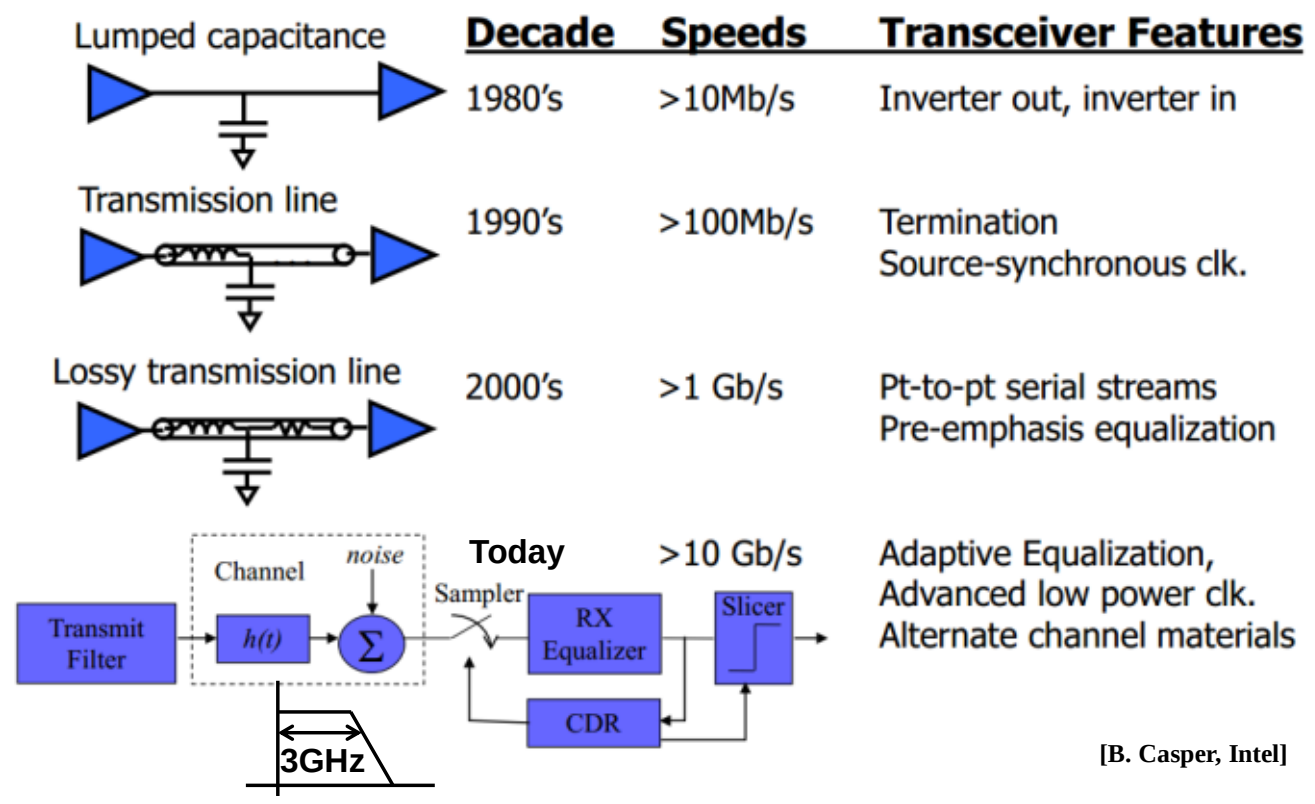

Fig. 3 Evolution of serial interface over the several decades

## III. STATE-OF-ART EQUALIZATION SOLUTIONS

Fig .4 illustrated the common interface data link transmitter (TX) and receiver (RX) architecture with a comprehensive combination of all the equalization mentioned above techniques.

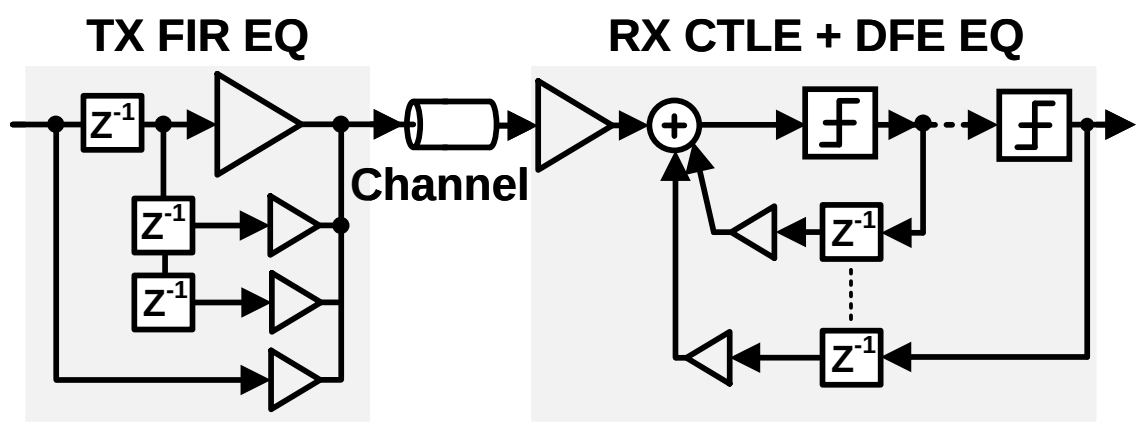

Fig. 4 Conventional comprehensive combination of equalization techniques

In the time-domain analysis, if a single bit is sent out onto the channel there will a long tail existing, which is very severe inter-symbol-interference (ISI). From the frequency-domain point of view, if the data bandwidth is lower than the available channel bandwidth, for example, 2Gb/s, there would not be any effect observed if channel 3-dB bandwidth is 3 GHz. However, if we are trying to send more than 10Gb/s data, eye diagram will completely close due to strong ISI and bit error rate (BER) will be awful. Similarly, for the low-cost connector and cable channel, there are a lot of discontinuities, such as bumps, vias, traces in the package, traces on the PCB, connector transition. All of these in the signal path might create strong resonances. They are also sensitive to fabrication variations. We can find for this particular low-cost cable; a single-bit transmission created more than 20 unit intervals (UI) after the channel at the receiver input. In this situation, if two or more bits are sent out onto the channel continuously, the receiver cannot tell it is one or zero without any equalization techniques. To make the matter worse, the location and depth of these frequency notches are very sensitive to PCB or connector fabrication variations. It is pretty challenging and power-hungry to use equalization technique to equalize so many non-ideal channel effects.

There are several equalization methods commonly used. They are FIR filter equalization at the transmitter side, CTLE (continuous time linear equalization) and nonlinear DFE (decision feedback equalization) at the receiver side. All of them have their advantages and disadvantages. The practical way is to combine all them together to achieve optimal operation point regarding energy efficiency, maximum data bandwidth.

## IV. CONCEPT OF RADIO FREQUENCY MODULATED SIGNALING

The fundamental consideration of multi-band signaling is precisely the same as the cable TV system or wireless orthogonal frequency-division multiplexing (OFDM) system. However, both cable TV and wireless OFDM system are relatively narrow band systems, while the serial interface is broadband. Channel conditions of serial interface are also very different.

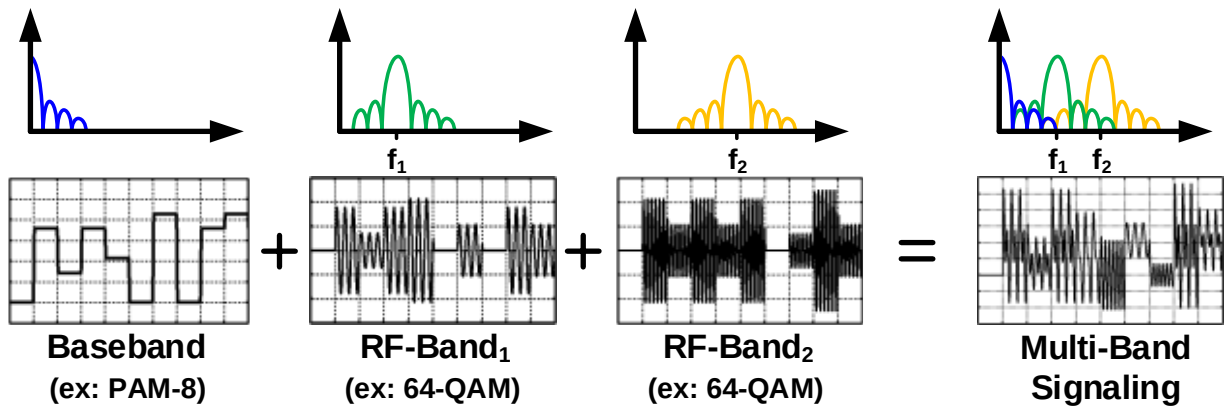

Fig. 5 Concept of radio frequency modulated signaling

In Fig. 5, PAM-8 and 64-QAM are shown as an example. 15 parallel data streams are running at 1 Gb/s as a data source. The PAM-8 modulator modulates three of them, the time-domain waveform of which are still in base-band but with multi-level features. Six of them pass to the 64-QAM modulator, the time-domain waveform of which is modulated by RF carrier frequency $f_1$. Similarly, another 6 of them are modulated by another RF carrier frequency $f_2$. Then, all of these waveforms are summed together. There are one baseband, one RF band at $f_1$ and another RF band at $f_2$ in the frequency domain,

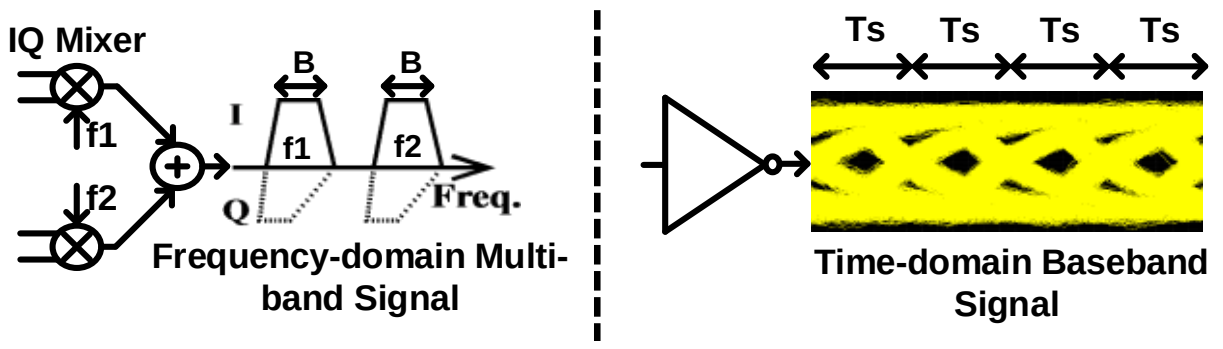

Fig. 6 Concept comparison of radio frequency modulated signaling and baseband signaling

respectively.

The merits of frequency-domain multi-band signaling over time-domain baseband are emphasized in Fig. 6. (1) Multi-band signaling can offer simultaneous and orthogonal communication channels in freq. domain; (2) It is easy to adapt with channel frequency notches by smartly choosing carrier frequency; (3) Multi-band signaling can relax equalization requirement because of self-equalization effect.

## V. RF-I TECHNOLOGY DEVELOPMENT OVERVIEW

The first (radio-frequency interconnect) RF-I transceiver was published in 2009 VLSI by Dr. Tam [27] as shown in Fig. 7. At that time, 30-GHz and 50-GHz radio frequencies (mm-wave frequencies) were used. It achieved 10Gb/s aggregated data rate. However, the channel is only 5mm on-chip transmission line for point-to-point communion. It is a practical data link for network-on-chip (NoC) and accelerator-rich architectures (ARA) applications. However, the channel distance is too short for high-performance distributed computing (HPDC) and cluster computing (CC) applications. The modulation scheme was non-coherent on-off key, which is the simplest non-coherent modulation scheme but also with the least power and hardware overhead.

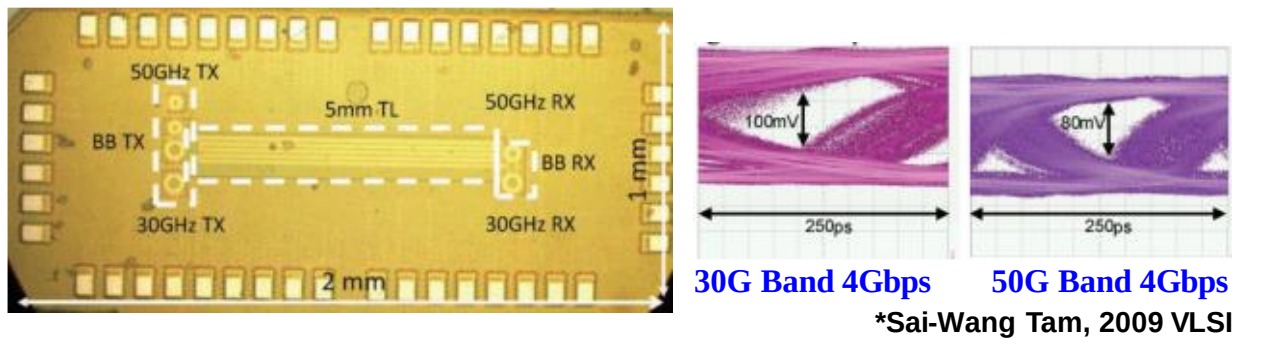

Fig. 7 On-chip RF-I transceiver in 2009 VLSI

The second version of RF-I system was demonstrated by Dr. Kim in 2012 ISSCC [28] as shown in Fig. 8. 18GHz RF carrier frequency was used. Moreover, it achieved 8Gb/s aggregated data rate. The channel condition was much more challenging, compared with the first generation RF-I. It included 5-cm PCB traces on the FR4 material. It was still point-to-point communication with on-off key modulation.

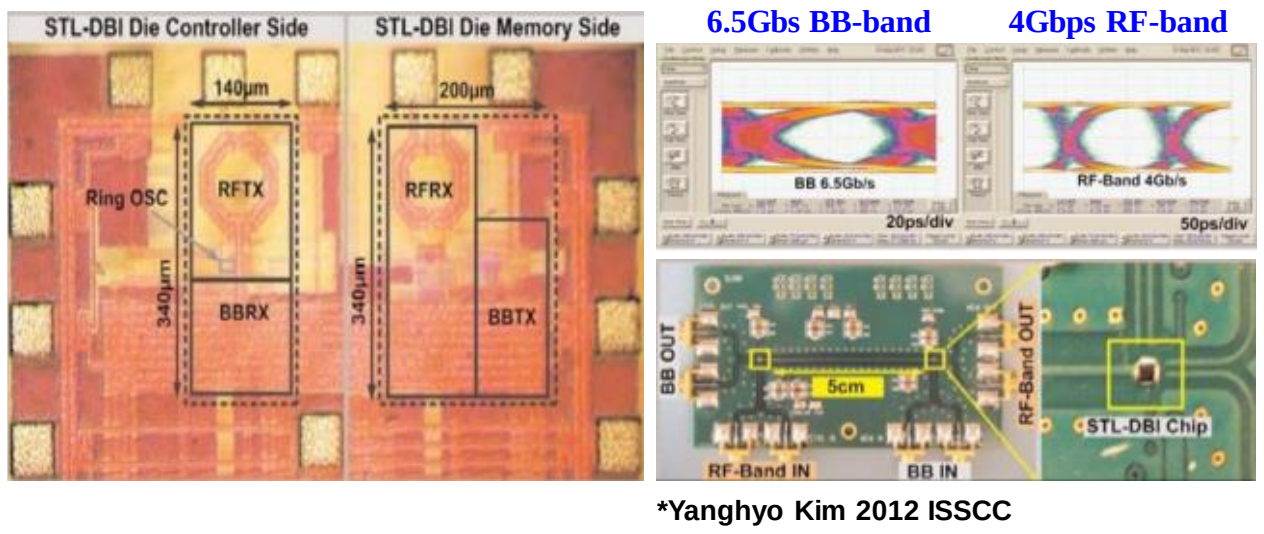

Fig. 8 On-board RF-I transceiver in 2012 ISSCC

A more advanced version of multi-band RF interconnects transceiver is realized [15] as Fig. 9 shown. It is proved that if one wants to extend communication distance and makes the whole serial interface more industrial friendly. The carrier frequencies have to be reduced from mm-wave frequency to within 10-GHz rather than using millimeter wave frequency range, in which range, it was difficult to achieve high energy efficiency due to skin effect metallic loss and dielectric material loss at high frequency. For this implementation, five carrier frequencies are used. It achieved 4 Gb/s aggregated data rate. Moreover, the channel is 2-inch transmission line on FR-4 PCB for point-to-point communication. It was the first time to demonstrate coherent modulation with QPSK in multi-band signaling serial interface transceiver.

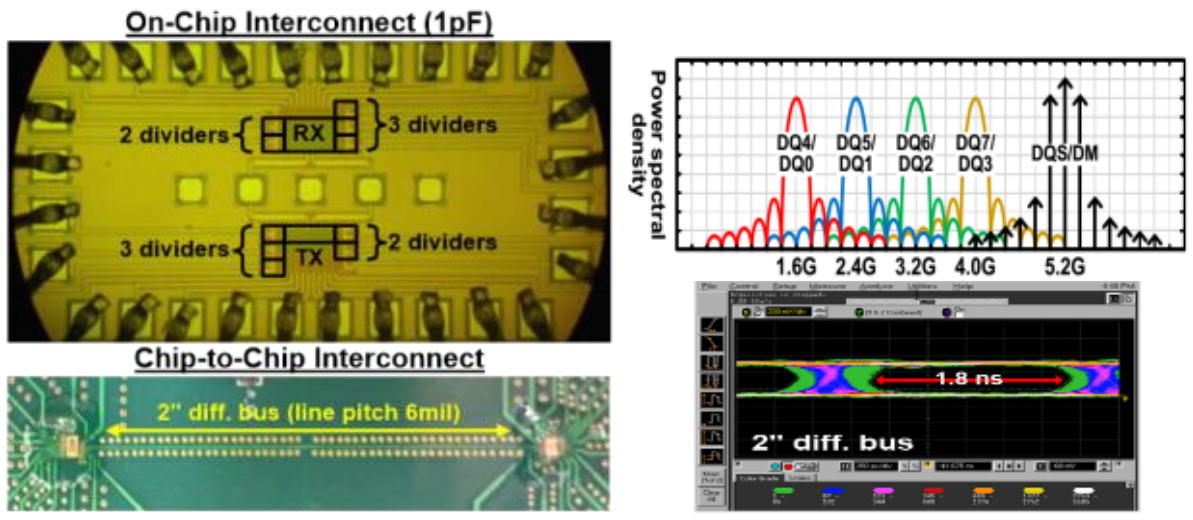

*W. H. Cho, *2015 CICC*

Fig. 9 On-board RF-I transceiver with QPSK in 2015 CICC

The updated version of multi-band RF serial interconnects transceiver uses 3GHz and 6GHz carrier frequencies to achieved 10 Gb/s aggregated data rate per differential pair [16], as shown in Fig. 10. The channel is 2-inch copper traces on FR-4 PCB. More significantly, both point-to-point communication and multi-drop channel communication are demonstrated with 16-QAM coherent modulation scheme.

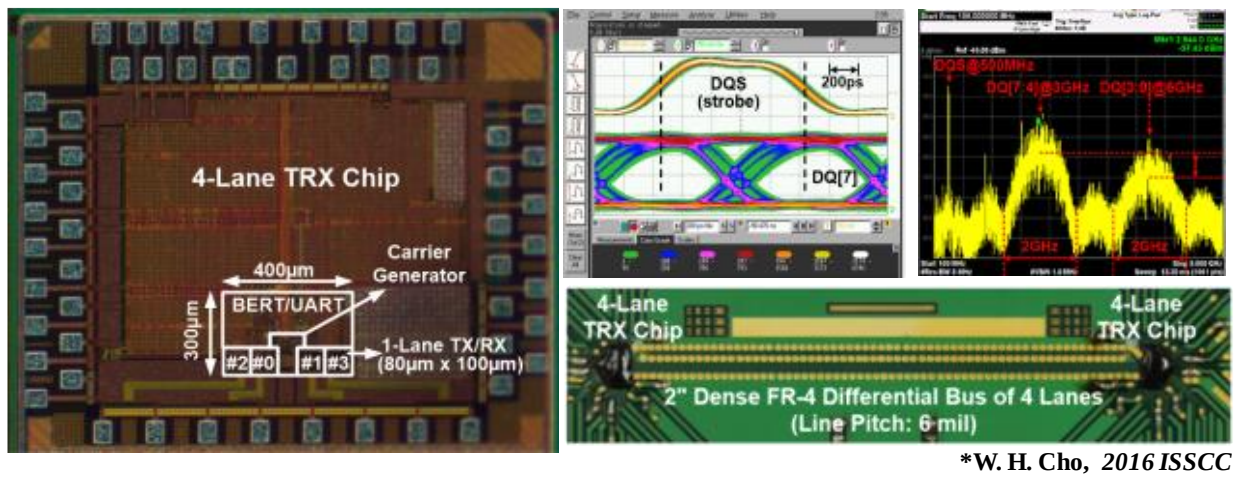

*W. H. Cho, *2016 ISSCC*

Fig. 10 On-board multi-band RF serial interconnect transceiver with 16-QAM supporting multi-drop bus (MDB) channel in 2016 ISSCC

The latest generation of RF-I transceiver also uses 3GHz and 6GHz carrier frequencies to achieved 16 Gb/s aggregated data rate per differential pair [17], as shown Fig. 11. The most advanced feature of this generation RF-I is cognitive to different data transmission channel conditions. It can learn tough channel conditions, such as multi-drop bus memory interface channel, and low-cost cable/connector channel. Simultaneously, it achieves less than one pJ/bit energy efficiency. The TX features learning an arbitrary channel response by sending a sweep of continuous wave, then detecting power level at receiver (RX) side, and accordingly adapting modulation scheme, data bandwidth and carrier frequency based on detected channel information. The supported modulation scheme ranges from NRZ/QPSK to PAM-16/256-QAM. The highly re-configurable TX is capable of dealing with low-cost serial channels, such as low-cost connectors, cables or multi-drop buses (MDB) with deep and narrow notches in the frequency domain (e.g., 40 dB loss at notches). The adaptive multi-band scheme mitigates equalization requirement and enhances the energy efficiency by avoiding frequency notches and utilizing the maximum available signal-to-noise ratio (SNR) and channel bandwidth.

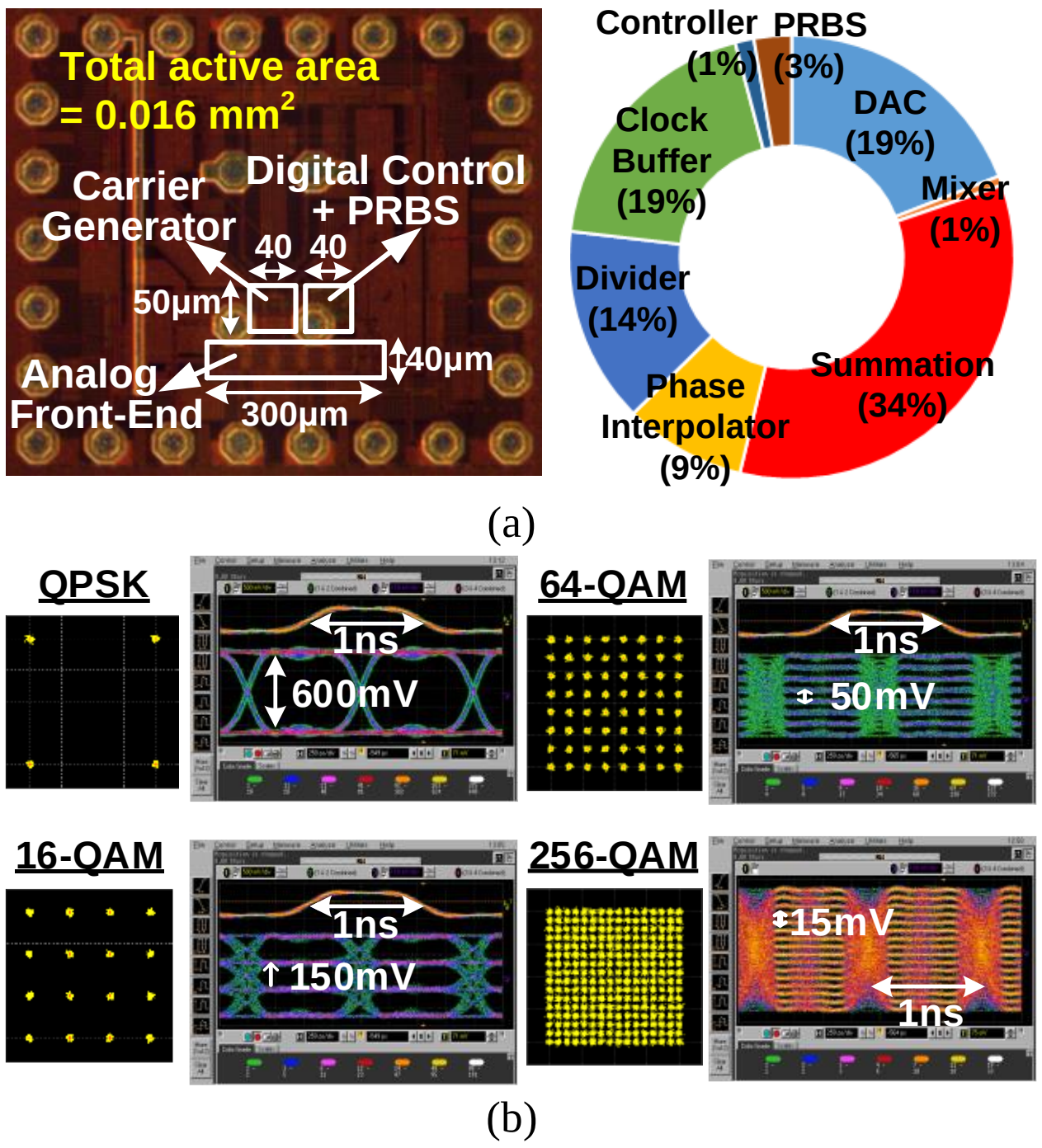

Fig. 11 Cognitive RF-I transmitter with reconfigurable coherent modulation up to 256-QAM in 2016 VLSI

TABLE I
DIFFERENT GENERATIONS OF RF-I COMPARISON

| Paper | Data Rate/diff. pair | Modulation Scheme | Channel Condition | Latency | Energy Efficiency |
|---|---|---|---|---|---|
| 2009 VLIS [27] | 8 Gb/s | OOK | on-chip trace | Low (wo/ clock forwarding) | 15 pJ/ bit |
| 2012 ISSCC [28] | 10 Gb/s | OOK | on-board trace | Low (wo/ clock forwarding) | 10 pJ/ bit |
| 2015 CICC [15] | 10 Gb/s | QPSK | on-board trace | Low (wo/ clock forwarding) | 2 pJ/ bit |
| 2016 ISSCC [16] | 10 Gb/s | 16-QAM / PAM-4 | on-board trace | Low (w/ clock forwarding) | 1 pJ/ bit |
| 2016 VLSI [17] | 16 Gb/s | 256-QAM / PAM-16 | on-board trace, MDB, low-cost cable | Low (w/ clock forwarding) | < 1 pJ/bit |

## VI. CONCLUSION

In this paper, the radio frequency modulation signaling interconnect is introduced, compared the conventional baseband signaling. We overview the different generations of radio-frequency interconnect (RF-I) technology. Table I compares the performance differences of different generations regarding data rate, modulation schemes, channel conditions, latency and energy efficiency.